\documentclass[a4paper,oneside,final,notitlepage,onecolumn,12pt]{article}
\usepackage{amsfonts}
\usepackage{epsf}
\usepackage{graphicx}
\usepackage{amssymb,eso-pic}
\usepackage{latexsym}
\usepackage{tabularx}
\usepackage{amsxtra} 
\usepackage{t1enc}
\usepackage{amsmath,accents}
\usepackage{auto-pst-pdf}
\usepackage{cancel}

\usepackage{ifpdf,epsfig,array,amsmath,amssymb}

\usepackage{psfrag}
 
\usepackage{authblk}
\usepackage{mathrsfs}
\usepackage{amsthm}

\usepackage{amsfonts}
\usepackage{epsf}
\usepackage{graphicx}
\usepackage{amssymb,eso-pic}
\usepackage{latexsym}
\usepackage{tabularx}
\usepackage{amsxtra} 
\usepackage{t1enc}
\usepackage{amsmath,accents}
\usepackage{auto-pst-pdf}
\usepackage{cancel}

\usepackage{ifpdf,epsfig,array,amsmath,amssymb}

\usepackage{psfrag}

\psfrag{x}{\footnotesize $x$}
\psfrag{y}{\footnotesize $y$}
\psfrag{z}{\footnotesize $z$} 

\psfrag{x=A}{\tiny $(A,0,0)$}
\psfrag{x=-A}{\tiny $(-A,0,0)$}
\psfrag{y=A}{\tiny $(0,A,0)$}
\psfrag{y=-A}{\tiny $(0,-A,0)$}
\psfrag{z=A}{\tiny $(0,0,A)$}
\psfrag{z=-A}{\tiny $(0,0,-A)$}

\psfrag{s1}{\footnotesize $\boldsymbol{\vec{s}}{}^{\hskip.2mm{}^{{}_{\,[1]}}}$}
\psfrag{v1}{\footnotesize $\boldsymbol{\vec{v}}{}^{\hskip.2mm{}^{{}_{[1]}}}$}
\psfrag{d1}{\footnotesize $\boldsymbol{\vec{d}}{}^{\hskip.2mm{}^{{}_{\,[1]}}}$}
\psfrag{s2}{\footnotesize $\boldsymbol{\vec{s}}{}^{\hskip.2mm{}^{{}_{\,[2]}}}$}
\psfrag{v2}{\footnotesize $\boldsymbol{\vec{v}}{}^{\hskip.2mm{}^{{}_{[2]}}}$}
\psfrag{d2}{\footnotesize $\boldsymbol{\vec{d}}{}^{\hskip.2mm{}^{{}_{\,[2]}}}$}
\psfrag{s3}{\footnotesize $\boldsymbol{\vec{s}}{}^{\hskip.2mm{}^{{}_{\,[3]}}}$}
\psfrag{v3}{\footnotesize $\boldsymbol{\vec{v}}{}^{\hskip.2mm{}^{{}_{[3]}}}$}
\psfrag{d3}{\footnotesize $\boldsymbol{\vec{d}}{}^{\hskip.2mm{}^{{}_{\,[3]}}}$}

\newcommand{\interior}[1]{\accentset{\smash{\raisebox{-0.12ex}{$\scriptstyle\circ$}}}{#1}\rule{0pt}{2.3ex}}
\newcommand{\instar}[1]{\accentset{\smash{\raisebox{-0.12ex}{$\scriptstyle\star$}}}{#1}\rule{0pt}{2.3ex}}
\makeatletter
\newcommand*{\bigcdot}{}
\DeclareRobustCommand*{\bigcdot}{%
	\mathbin{\mathpalette\bigcdot@{}}%
}
\newcommand*{\bigcdot@scalefactor}{.5}
\newcommand*{\bigcdot@widthfactor}{1.15}
\newcommand*{\bigcdot@}[2]{%
	\sbox0{$#1\vcenter{}$}
	\sbox2{$#1\cdot\m@th$}%
	\hbox to \bigcdot@widthfactor\wd2{%
		\hfill
		\raise\ht0\hbox{%
			\scalebox{\bigcdot@scalefactor}{%
				\lower\ht0\hbox{$#1\bullet\m@th$}%
			}%
		}%
		\hfil
	}%
}
\makeatother

\newcommand{\iiindot}[1]{\accentset{\smash{\raisebox{-0.1ex}{$\scriptstyle{\bigcdot}$}}}{#1}\rule{0pt}{2.3ex}}
\def\sqint{{\scriptstyle\square}\hskip-.321cm\int}
\def\gsqint{{\square}\hskip-.465cm\int}
\def\preA{\hskip-0.5mm{}^{{}_{}^{(A)}}\hskip-0.5mm}

\usepackage[normalem]{ulem}
\usepackage{color}
\definecolor{blue}{rgb}{0,0,1}
\definecolor{red}{rgb}{1,0,0}

\makeatother

\DeclareFontFamily{OT1}{rsfs}{} \DeclareFontShape{OT1}{rsfs}{m}{n}{
	<-7> rsfs5 <7-10> rsfs7 <10-> rsfs10}{}
\DeclareMathAlphabet{\mathscr}{OT1}{rsfs}{m}{n}

\def\sc{{\hskip 3.5pt {{}^{{}^{{}_{{}_{\bowtie}}}}} \kern -8.pt{}}}  
\def\SC{{\hskip 3.5pt {{}^{{}^{{}^{{}_{{}_{\bowtie}}}}}} \kern -10.5pt{}}}

\newtheorem{theorem}{Theorem}

\newtheorem*{example*}{Example}
\newtheorem*{condition*}{Condition}
 
\usepackage[normalem]{ulem}
\usepackage{color}

\newcounter{mnotecount}

\newcommand{\mnotex}[1]
{\protect{\stepcounter{mnotecount}}$^{\mbox{\footnotesize $\bullet$\themnotecount}}$ 
	\marginpar{\color{red}
		\raggedright\tiny\em
		$\!\!\!\!\!\!\,\bullet$\themnotecount: #1} }

\begin{document}

\vskip-3cm
	\title{Can we prescribe the physical parameters of multiple black holes?
	}
	
	\author{{Istv\'an R\'acz}\,\thanks{ ~email: racz.istvan@wigner.hu}}  
	
	\affil{Wigner RCP, H-1121 Budapest, Konkoly Thege Mikl\'{o}s \'{u}t  29-33, Hungary}	
	
	\maketitle

\begin{abstract}
	
The parabolic-hyperbolic form of the constraints and superposed Kerr-Schild black holes have already been used to provide a radically new initialization of binary black hole configurations. The method generalizes straightforwardly to multiple black hole systems. This paper is to verify that each of the global  Arnowitt-Deser-Misner quantities of the constructed multiple black hole initial data can always be prescribed, as desired, in advance of solving the constraints.  These global charges are shown to be uniquely determined by the physical parameters of the involved individual Kerr-Schild black holes.
 
\end{abstract} 


\section{Introduction}
\label{introduction}
\setcounter{equation}{0}

Binary black holes are considered to be the foremost vital sources for the emerging field of gravitational-wave astrophysics. Multiple black hole systems, above obvious curiosities, may also serve as natural gravitational waves sources. Investigation of the dynamics of these systems starts with a careful initialization. 
This may be done by applying the elliptic method \cite{lich, york0} (see also \cite{choquet, ch-co-is}) or either of the evolutionary form of the constraints introduced in \cite{racz_constraints}.

\medskip

In \cite{racz_bh_ID,racz_bh_ID-suppl}, the parabolic-hyperbolic formulation of Hamiltonian and momentum constraints, along with superposing individual Kerr-Schild black holes, was applied to construct initial data for binary black hole configurations. The only technical restriction was that each of the initial speeds is parallel to and each of the spin vectors is orthogonal to a plane of some background Euclidean space, respectively. For this class, the existence and uniqueness of (at least) $C^2$ solutions to the parabolic-hyperbolic form of the constraints is outlined in \cite{racz_bh_ID,racz_bh_ID-suppl}. 
This new method has also been successfully applied to determine initial data numerically for individual and binary black hole systems \cite{yorgos, Anna}. Notably, the very same construction can also be used to initialize multiple black hole systems. 
There are no restrictions on the masses, speeds, spins, and distances of the individual black holes; thereby, this set contains many physically realistic initial data configurations.

\medskip 

This paper will focus on the Arnowitt-Deser-Misner (ADM) charges of multiple black hole systems. Accordingly, superposing individual Kerr-Schild black holes, each located momentarily on a plane in some background Euclidean space with speeds parallel to and spins orthogonal to the distinguished plane. By adopting constructive elements of the proposal in \cite{racz_bh_ID, racz_bh_ID-suppl}, we shall choose the free data to the initial-boundary value problem, derived from the parabolic-hyperbolic form of the constraints, using superposed Kerr-Schild black holes. Clearly, if the individual Kerr-Schild black holes are widely separated the initial data, satisfying the constraints, will only slightly differ from the superposed data induced on a $t=const$ time-slice. Therefore, it is highly plausible that the craved global solutions exist. Note, however, that the verification of the primary result of the present paper refers only to the specific choice of the free data, and it does not require detailed knowledge of solutions. Therefore, no attempt will be made to deal with the global existence and uniqueness of solutions to the aforementioned initial-boundary value problem. Instead, assuming that appropriate free data has been chosen, we will assume that the global existence of asymptotically flat solutions to the aforementioned initial-boundary value problem to hold in an analogy of the arguments applied in \cite{racz_bh_ID, racz_bh_ID-suppl}.

\medskip 
Note that the metric (\ref{eq:bks4}) of superposed Kerr-Schild black holes---though not satisfying Einstein's equations---is asymptotically flat \cite{racz_bh_ID}. Therefore, it is plausible that solutions to the specific initial-boundary value problem will also be asymptotically flat. The asymptotic form of the metric (\ref{eq:bks4}), along with the geometric assumptions imposed in our construction, guarantee that well-defined ADM mass, center of mass, linear and angular momenta can always be associated with the corresponding multiple black hole system.   
The main point in this paper is that all the ADM quantities of multiple black hole systems are determined by the rest masses, positions, velocities, and spins of the involved individual black holes.

\medskip 

This result immediately raises the question if any other formulation of the constraints can provide an
analogous determination of the ADM quantities. For instance, the method proposed by Bowen and York \cite{bowen-york}, in principle, allows to prescribe the ADM linear and angular momenta by solving the momentum constraint explicitly \cite{bs, bowen-york}. However, to do so, they had to apply a restricted set of basic variables. In particular, to guarantee that the Hamiltonian and momentum constraints decouple, the authors had to assume vanishing of the mean curvature, in addition to assuming the conformal flatness of the Riemannian metric $h_{ij}$. One of the unfavorable consequences of these technical assumptions is that they are known to be so strong that they exclude even the Kerr black hole solution from the outset \cite{garat_price, kroon}. One should also mention here that, within the setup proposed by Bowen and York \cite{bowen-york}, there is no way to get an analogous control on the ADM mass or the center of mass.

\medskip 

In the context of the determinacy of the ADM quantities, one should also mention the construction applied in \cite{ch-co-is}, where, by combining the gluing techniques with Kerr-Schild black holes, an interesting initialization of multiple black hole systems was proposed. Indeed, as Kerr-Schild black holes were applied in \cite{ch-co-is} and our proposal also rests upon using these types of black holes, one would expect that analogous determinacy of the ADM quantities applies to both of these approaches. It is, however, not the case, as gluing requires the use of the elliptic method that starts by a conformal rescaling of the basic variables. In turn, gluing gets somewhat implicit, which does not allow---apart from the extreme case with infinitely separated individual black holes---to have complete control on the ADM quantities \cite{ch-co-is}. Yet another unfavorable consequence of using the conformal method is that intermediate regions---where the gluing happens---have to be allocated to each of the involved Kerr-Schild black holes. This, however, does not allow to set the initial distances of these black holes to be arbitrary, and, as stated explicitly in \cite{ch-co-is}, ``they must be separated by a distance above a certain threshold''.

\medskip

It is important to emphasize that our proposal does not impose analogous restrictions on the distances of individual black holes. Yet, it provides an unprecedented complete control on the ADM parameters of multiple black hole systems. As this happens in advance of solving the constraints, an unprecedented fine-tuning of the complete set of ADM parameters of the to-be solutions is possible.

\medskip 

This paper is structured as follows. In Section \ref{sec: prelim}, first a brief account on the parabolic-hyperbolic form of the constraints is given. This is followed by recalling the notion of asymptotic flatness, the definition of the ADM quantities and the superposition of Kerr-Schild black holes in subsections \ref{asymptotic_flatness_adm_quantities} and \ref{superposed}. In Section \ref{initial-boundary}, the choice of the freely specifiable variables and the initial-boundary data, applied in determining multiple black hole initial data, and the pertinent fall-off properties are discussed. Section \ref{adm_quantities} is to present our mane result, containing a case by case verification of the statement that the ADM quantities of the superposed Kerr-Schild black holes and the corresponding multiple black hole initial data are pairwise equal to each other. The paper is closed, in Section \ref{sec: conclusions}, by our final remarks.

\section{Preliminaries}\label{sec: prelim}
\setcounter{equation}{0}

Initial data relevant for the vacuum Einstein's equations is comprised of a Riemannian metric $h_{ij}$ and a symmetric tensor field $K_{ij}$. Both of these fields are assumed to be given on a three-dimensional manifold $\Sigma$. They are not arbitrary as they have to satisfy the constraints which read as (see, e.g.~\cite{choquet})
\begin{align} 
	{}^{{}^{(3)}}\hskip-1mm R + \left({K^{j}}_{j}\right)^2 - K_{ij} K^{ij}=0 \label{expl_eh}\\
	D_j {K^{j}}_{i} - D_i {K^{j}}_{j}=0\,,\label{expl_em}
\end{align}
where ${}^{{}^{(3)}}\hskip-1mm R$ and $D_i$ denote the scalar curvature and the covariant derivative operator associated with $h_{ij}$, respectively. 

\subsection{The parabolic-hyperbolic form of constraints}

The essential steps in deriving the parabolic-hyperbolic form of the constraints are as follows:
Assume, for simplicity, that there exists a smooth function  $\rho: \Sigma\rightarrow \mathbb{R}$ such that the $\rho=const$ surfaces (denoted also by $\mathscr{S}_\rho$) provide a foliation of $\Sigma$. We assume that the $\mathscr{S}_\rho$ surfaces are homologous to each other and the transversal one form $D_i\rho$ to these level surfaces does not vanish on $\Sigma$. The unite normal $\widehat n_i$ to the $\mathscr{S}_\rho$ surfaces is given then as 
\begin{equation}
\widehat n_i = \big[h^{ef}(D_e\rho) (D_f\rho)  \big]^{-\tfrac12}\,D_i\rho\,.
\end{equation}
Choosing then a vector field $\rho^i$ on $\Sigma$ such that $\rho^i \partial_i \rho=1$, and considering its parallel and orthogonal parts we get
\begin{equation}\label{nhat}
\rho^i=\widehat{N}\,\widehat n^i+{\widehat N}{}^i\,,
\end{equation}
where $\widehat N$ and $\widehat N^i$ stand for the lapse and shift of $\rho^i$, respectively, and $\widehat n{}^i=h^{ij}\,\widehat n_j$. 


\medskip

Analogous decomposition of the metric $h_{ij}$ and the symmetric tensor field $K_{ij}$ gives
\begin{equation}\label{hij}
h_{ij}=\widehat \gamma_{ij}+\widehat  n_i \widehat n_j \,, \quad {\rm and} \quad  
K_{ij}= \boldsymbol\kappa \,\widehat n_i \widehat n_j  + \left[\widehat n_i \,{\rm\bf k}{}_j  
+ \widehat n_j\,{\rm\bf k}{}_i\right]  + {\rm\bf K}_{ij}\,,
\end{equation}
where $\widehat \gamma_{ij}=\widehat \gamma{}^e{}_i\widehat \gamma{}^f{}_jh_{ef}$ is the induced metric on the $\rho=const$ level surfaces, and
\begin{equation}
\boldsymbol\kappa= \widehat n^k\widehat  n^l\,K_{kl}\,,\ {\rm\bf k}{}_{i} = {\widehat \gamma}^{k}{}_{i} \,\widehat  n^l\, K_{kl}\,,\ {\rm\bf K}_{ij} = {\widehat \gamma}^{k}{}_{i} {\widehat \gamma}^{l}{}_{j}\,K_{kl}\,, 
\end{equation}
where ${\widehat \gamma}^{k}{}_{i}$ denotes the projection operator 
\begin{equation}
	{\widehat \gamma}^{k}{}_{i} = h^k{}_i- \widehat n{}^k\widehat n_i\,.
\end{equation}
It is also rewarding to introduce the trace and trace-free part of ${\rm\bf K}_{ij}$ defined as
\begin{equation}\label{trace_free}
{\rm\bf K}^l{}_{l}=\widehat\gamma^{kl}\,{\rm\bf K}_{kl} \quad {\rm and} \quad  \interior{\rm\bf K}_{ij}={\rm\bf K}_{ij}-\tfrac12\,\widehat \gamma_{ij}\,{\rm\bf K}^l{}_{l}\,.
\end{equation}

\medskip

Using the new variables $\widehat N,\widehat N^i,\widehat \gamma_{ij}, \boldsymbol\kappa,{\rm\bf k}{}_{i}, \interior{\rm\bf K}_{ij},{\rm\bf K}^l{}_{l}$ the constraints can be seen to be equivalent to the parabolic-hyperbolic system for $\widehat N,{\rm\bf k}{}_{i}$ and ${\rm\bf K}^l{}_{l}$ \cite{racz_constraints} 
\begin{align}
	{}& \instar{K}\,[\,(\partial_{\rho} \widehat N) - \widehat N{}^l(\widehat D_l\widehat N) \,] = \widehat N^{2} (\widehat D^l \widehat D_l \widehat N) + \mathcal{A}\,\widehat N + \mathcal{B}\,\widehat N{}^{3} \,, \label{bern_pde} \\ 
	{}& \mathscr{L}_{\widehat n} {\rm\bf k}{}_{i} - \tfrac12\,\widehat D_i ({\rm\bf K}^l{}_{l}) - \widehat D_i\boldsymbol\kappa + \widehat D^l \interior{\rm\bf K}{}_{li} + \widehat N \instar{K}\,{\rm\bf k}{}_{i}  + [\,\boldsymbol\kappa-\tfrac12\, ({\rm\bf K}^l{}_{l})\,]\,\iiindot{\widehat n}{}_i - \iiindot{\widehat n}{}^l\,\interior{\rm\bf K}_{li} = 0 \label{par_const_n} \\
	{}& \mathscr{L}_{\widehat n}({\rm\bf K}^l{}_{l}) - \widehat D^l {\rm\bf k}_{l} - \widehat N \instar{K}\,[\,\boldsymbol\kappa-\tfrac12\, ({\rm\bf K}^l{}_{l})\,]  + \widehat N\,\interior{\rm\bf K}{}_{kl}\instar{K}{}^{kl}  + 2\,\iiindot{\widehat n}{}^l\, {\rm\bf k}_{l}  = 0\,. \label{ort_const_n}
\end{align}
Here, $\widehat D_i$ denotes the covariant derivative operator associated with $\widehat \gamma_{ij}$  and
\begin{align}
\instar{K}_{ij}= {}& \tfrac12\mathscr{L}_{\rho} {\widehat \gamma}_{ij} -\widehat  D_{(i}\widehat N_{j)} \,,\label{hatext} \\ 
 \instar{K}  = {}& \tfrac12\,{\widehat \gamma}^{ij}\mathscr{L}_{\rho} {\widehat \gamma}_{ij} -  \widehat D_j\widehat N^j\,,\label{trhatext} \\
\mathcal{A} = {}& (\partial_{\rho} \instar{K}) - \widehat N{}^l (\widehat D_l \instar{K}) + \tfrac{1}{2}[\,\instar{K}^2 + \instar{K}{}_{kl} \instar{K}{}^{kl}\,] \label{A}  \\
\mathcal{B} = {}& - \tfrac12\,\bigl[\widehat R + 2\,\boldsymbol\kappa\,({\rm\bf K}^l{}_{l})+\tfrac12\,({\rm\bf K}^l{}_{l})^2 
-2\,{\rm\bf k}{}^{l}{\rm\bf k}{}_{l}  - \interior{\rm\bf K}{}_{kl}\,\interior{\rm\bf K}{}^{kl}\,\bigr] \label{B}
\\ \iiindot{\widehat n}{}_k = {}& {\widehat n}{}^lD_l{\widehat n}{}_k=-{\widehat D}_k(\ln{\widehat N})\,.
\end{align}

\medskip

The variables 
$\widehat N^i,\widehat \gamma_{ij}, \boldsymbol\kappa$ and $\interior{\rm\bf K}_{ij}$ are unconstrained whence they are freely specifiable throughout $\Sigma$. The well-posedness of the coupled system (\ref{bern_pde})--(\ref{ort_const_n}) is guaranteed if (\ref{bern_pde}) is uniformly parabolic. It was shown in \cite{racz_constraints} that this happens in those subregions of $\Sigma$ where $\instar{K}$ is either strictly positive or negative. Note also that, as $\instar{K}$ depends on the freely specifiable fields $\widehat \gamma_{ij}$ and $\widehat N^i$ exclusively, the sign of $\instar{K}$ (at least locally) is adjustable according to the needs of specific problems to be solved \cite{racz_constraints,racz_bh_ID}.

\medskip

Note that, in addition to the freely specifiable variables, on one of the $\rho=const$ level surfaces initial data has also to be chosen for the constrained variables \cite{racz_constraints}. Once this has been done, i.e.\;smooth data has been chosen for $\widehat N, {\rm\bf k}{}_{i}$ and ${\rm\bf K}^l{}_{l}$, then a unique smooth solution exists to (\ref{bern_pde})--(\ref{ort_const_n}) in the domain of dependence of that $\rho=const$ level surface. This, in general, can be seen to be global as the hyperbolic part of the system,  see (\ref{par_const_n}) and (\ref{ort_const_n}), is linear in ${\rm\bf k}{}_{i}$ and ${\rm\bf K}^l{}_{l}$. It is also important that the fields $h_{ij}$ and $K_{ij}$ that can be reconstructed from such a solution, and from the freely specifiable fields, do satisfy the Hamiltonian and momentum constraints (\ref{expl_eh}) and (\ref{expl_em}). 

\subsection{Asymptotic flattness and the ADM quantities}
\label{asymptotic_flatness_adm_quantities}

Before turning to the specific class of solutions to the parabolic-hyperbolic system (\ref{bern_pde})--(\ref{ort_const_n}), it is rewarding to have a glance at the generic notion of asymptotic flatness, along with the conditions ensuring the existence of well-defined ADM charges.

\medskip

Our model is based on the use of the superposed Kerr-Schild metric, whence, the singularities of the multiple black hole system will be arranged to be located in a finite ball $\mathscr{B}$ in $\mathbb{R}^3$. Therefore it suffices to assume the existence of a single asymptotically flat end. (See section \ref{initial-boundary} below for further specifications.) 

\medskip

The initial data set $(\Sigma, h_{ij}, K_{ij})$ is called strictly\,\footnote{Note that in Section \ref{adm_quantities}---likewise in many  other analogous investigations (see, e.g.\,\cite{ch})---weaker fall-off conditions could also be used. Nevertheless, hereafter, for convenience, and for definiteness, the above recalled stronger fall-off conditions  will be applied.} asymptotically Euclidean \cite{ch-co-is}, to order $\ell$, if in exterior to $\mathscr{B}$ i.e. in $\Sigma\setminus \mathscr{B}$, admissible asymptotically flat coordinates $x_i=(x,y,z)$ exist such that 
%
\begin{equation} \label{af-l2}
	\big|\,\partial^{\alpha}\big(\,h_{ij}-\delta_{ij}\,\big)(\vec{x})\,\big|=\mathcal{O}(\,|\vec{x}|^{-|\alpha|-1}\,)
	, \quad
	\big|\,\partial^{\beta}K_{ij}(\vec{x})\,\big|=\mathcal{O}(\,|\vec{x}|^{-|\beta|-2}\,),
\end{equation}
hold, where $|\vec{x}|=\sqrt{x_1^2+x_2^2+x_3^2}$, and where $\partial^\alpha$, with multi index $\alpha=\alpha_1+\alpha_2+\alpha_3$, stands for the composition of partial derivative operators $\partial_{x_1}^{\,\alpha_1}\partial_{x_2}^{\,\alpha_2}\partial_{x_3}^{\,\alpha_3}$, and for the multi indices $\alpha$ and $\beta$, for some value of $\ell$, the inequalities $|\alpha|\leq \ell+1$ and $|\beta|\leq \ell$ hold. For the arguments applied in this paper $\ell\geq 1$ will suffice. In most cases the operators $\partial_{x_i}$ will also be abbreviated as $\partial_{i}$.

\bigskip

It is known, that conditions in (\ref{af-l2}) can only guarantee the existence and finiteness of the four-momentum, and to have, in addition, well-defined center of mass and angular momentum the so-called Regge-Teitelboim asymptotic parity conditions need to be used which, in admissible coordinates, read as 
\begin{equation} \label{R-T}
\big|\,\partial^{\alpha}\big[\,h_{ij}(\vec{x})-h_{ij}(-\vec{x})\,\big]\,\big|=\mathcal{O}(\,|\vec{x}|^{-|\alpha|-2}\,)
, \quad
\big|\,\partial^{\beta}\big[\,K_{ij}(\vec{x})-K_{ij}(-\vec{x})\,\big]\,\big|=\mathcal{O}(\,|\vec{x}|^{-|\beta|-3}\,)\,.
\end{equation}
%

Assuming that both the asymptotic flatness and the Regge-Teitelboim conditions hold, the ADM mass, center of mass, linear and angular momenta are given by the flux integrals \cite{ch-co-is}
\begin{align}
	M^{\,ADM}= {} & \frac{1}{16\pi}\oint_\infty 
	\left[\, \partial_i h_{ij} - \partial_j h_{ii} \,\right] n^j  {\rm d}S  \label{int:Madm} \\	
	M^{\,ADM}d_i= {} &  \frac{1}{16\pi}\oint_\infty 
	\Big\{ x_i \left[\, \partial_k h_{kj} - \partial_j h_{kk} \,\right]   -  \left[\, h_{kj}\,\delta^k{}_i -  h_{kk}\,\delta_{ij} \,\right]
	\Big\}\, n^j   {\rm d}S \label{int:center:adm} \\
	{P}_i^{\,ADM}= {} & \frac{1}{8\pi}\oint_\infty  
	\left[\, K_{ij} - h_{kj}\,K^{l}{}_{l} \,\right] n^j  {\rm d}S \label{int:momentum:adm} \\
	{J}_i^{\,ADM}= {} & \frac{1}{8\pi} \oint_\infty 
	\left[\, K_{kj} - h_{kj}\,K^{l}{}_{l} \,\right]\, \epsilon_{i}{}^{lk}x_l\, n^j {\rm d}S \label{int:angmomentum:adm}\,,
\end{align}
where the symbol $\oint_\infty$ is meant to denote limits of integrals over spheres while their radii tend to infinity, whereas $n^i$ and $ {\rm d}S$ denote  the outward pointing unit normal and the volume element of the individual spheres of the sequences, respectively. Note that in (\ref{int:angmomentum:adm}) $\epsilon_{i}{}^{jk}x_j$ stands for the components of the three rotational Killing vector fields, defined with respect to the applied admissible asymptotically Euclidean coordinates $x_i$. 

\subsection{Superposed Kerr-Schild black holes}
\label{superposed}

The Kerr solution \cite{ks1} is known to take the Kerr-Schild form given as
\begin{equation}\label{eq:ksm}
	g_{\alpha\beta}=\eta_{\alpha\beta}+2 H \ell_{\alpha} \ell_{\beta}\,, 
\end{equation}
where
\begin{equation}\label{H-ell-kerr}
H=\frac{r^3M}{r^4+{a^2z^2}} \quad {\rm and} \quad \ell_{\alpha}=\left(1, \frac{r\,x+a\,y}{r^2+a^2},
\frac{r\,y-a\,x}{r^2+a^2},\frac{z}{r}  \right)\,,
\end{equation}
and where the Boyer-Lindquist radial coordinate $r$ is related to 
the spatial part of the inertial coordinates $x^\alpha=(t,x,y,z)$---which are asymptotically flat admissible coordinates---via the implicit relation 
\begin{equation}\label{imp-r-def}
\frac{x^2+y^2}{r^2+a^2}+\frac{z^2}{r^2}=1\,.
\end{equation}

It is well-known that generic displaced, boosted and spinning black holes can be produced by performing suitable Poincar\'e transformations on a Kerr black hole. It is also important that the Kerr-Schild metric is form-invariant under these transformations. In particular, even if a Lorentz transformation $x'{}^{\alpha}=\Lambda^{\alpha}{}_{\beta}\,x^{\beta}$ is performed 
the metric $g'_{\alpha\beta}$ in the new coordinates will retain the Kerr-Schild form $g'_{\alpha\beta}=\eta_{\alpha\beta}+2 H' \ell'_{\alpha} \ell'_{\beta}$, where
\begin{equation}
H'(x'{}^{\alpha})=H\left([\Lambda^{\alpha}{}_{\beta}]^{-1}x'{}^{\beta}\right) \quad {\rm and} \quad \ell'_{\beta}(x'{}^{\varepsilon})=\Lambda^{\alpha}{}_{\beta}\, \ell_{\alpha}\left([\Lambda^\varepsilon{}_\varphi]^{-1}x'{}^\varphi\right)\,.
\end{equation}

\medskip

As boosts and spatial rotations are special Lorentz transformations it is straightforward to construct models of moving black holes with preferably oriented speed and spin by performing suitable combinations of boosts and rotations on a Kerr black hole that is in rest and suitably oriented with respect to a auxiliary Minkowski background. Note also that displacement of these boosted and spinning black holes may be represented by a straightforward change in the argument of $H$ and $\ell_{\alpha}$ whence all the displaced, boosted and spinning individual black holes may be produced by applying the indicated transformations. 

\medskip

Now we are almost ready to combine the parabolic-hyperbolic form of the constraints with superposed Kerr-Schild black holes. As indicated in the introduction solving the constraints in their parabolic-hyperbolic form (\ref{bern_pde})--(\ref{ort_const_n}) requires  specifications of the unconstrained variables $\widehat N^i,\widehat \gamma_{ij}, \boldsymbol\kappa$ and $\interior{\rm\bf K}_{ij}$ everywhere on $\Sigma$, and, in addition, an 
initialization of constrained variables $\widehat N, {\rm\bf k}{}_{i}$ and ${\rm\bf K}^l{}_{l}$ on one of the $\rho=const$ level surfaces.

\medskip

In virtue of the results in \cite{racz_bh_ID} it is rewarding to start with the auxiliary metric
\begin{equation}\label{eq:bks4}
g_{\alpha\beta}= \eta_{\alpha\beta} + \sum_{I=1}^{N}\, 2\, H {}^{[I]}\ell_{\alpha} {}^{[I]}\ell_{\beta}{}^{[I]} 
\end{equation} 
yielded by superposing the contributions of individual black holes represented by the $(H {}^{[I]},  \ell_\alpha {}^{[I]})$, $I=1,\dots,N$,  pairs. This metric retains much of the algebraic simplicity of (\ref{eq:ksm}). For instance, the vector fields $\ell^{\alpha}{}^{[I]}$ remain null with respect to the background Minkowski metric, and they satisfy the geodesic equation $\ell^{\beta}{}^{[I]} \partial_{\beta} \, \ell^{\alpha}{}^{[I]}=0$, the functions $H {}^{[I]}$ satisfy the background wave equation $\eta^{\alpha\beta}\partial_{\alpha} \partial_{\beta} H{}^{[I]} =0$ (see, e.g.\,\cite{i_jeff}).

\medskip

From now on, to distinguish the real physical quantities from the auxiliary ones deduced from superimposed Kerr-Schild form (\ref{eq:bks4}), the latter will be labeled by the ``pre upper index'' ${}^{{}_{}^{(A)}}$. For instance,  ${{}^{{}_{}^{(A)}}\hskip-0.7mm} h_{ij}$ will stand for the three-metric 
\begin{equation}\label{eq:bks3}
\preA\hskip-0.2mm h_{ij}= \delta_{ij} + \sum_{I=1}^{N} \,2\, H {}^{[I]}\ell_i {}^{[I]}\ell_j{}^{[I]} 
\end{equation}
induced by the superposed Kerr-Schild form (\ref{eq:bks4}) on $t=const$ hypersurfaces, where $t$ is the time coordinate of the inertial system $x^\alpha=(t,x,y,z)$. Clearly, such a $t=const$ hypersurface may be assumed to be a Kerr-Schild time slice for each of the individual black holes, which implies that topologically it is simply the complement of the individual ``ring'' singularities in $\mathbb{R}^3$ as it was indicated in section\,\ref{asymptotic_flatness_adm_quantities}.

\medskip

Note that the metric (\ref{eq:bks4}) is not a solution to Einstein's equations yet it is asymptotically flat. This, in particular, means that the integrability of the scalar curvature of  \,$\preA\hskip-0.2mm h_{ij}$  (necessary to have,  for instance, well-defined ADM mass \cite{bartnik}) is guaranteed, and that  conditions in (\ref{af-l2}) and (\ref{R-T}) hold for the metric in (\ref{eq:bks4}), and, in turn, the superposed Kerr-Schild black holes can always be assigned with well-defined ADM charges. These quantities can be determined either by evaluating the flux integrals given in (\ref{int:Madm})--(\ref{int:angmomentum:adm}) or by taking into account the Poincar\'e transformations associated with displacements, boosts and rotations performed on the individual black holes. In either way we get the remarkably simple relations
\begin{align}
	\hskip-.3cm M^{\,ADM} = {} & \sum_{I=1}^{N} \, \gamma{}^{[I]} M{}^{[I]} \,, \label{M_adm_KS} \\ 
	M^{\,ADM}\vec{d} = {} & \sum_{I=1}^{N} \, \gamma{}^{[I]} M{}^{[I]} \vec{d}{ }^{\,\,[I]} \label{coM_adm_KS}\\
	\vec{P}^{\,ADM} = {} & \sum_{I=1}^{N} \, \gamma{}^{[I]} M{}^{[I]} \vec{v}{ }^{\,[I]}  \,,\label{P_adm_KS} \\ 
	\vec{J}^{\,ADM} =  {} & \sum_{I=1}^{N} \, \gamma{}^{[I]} \left\{M{}^{[I]} \vec{d}{}^{\,\,[I]} \hskip-0.1cm\times \vec{v}{}^{\,[I]}  + M{}^{[I]} {a}{}^{[I]}\vec{s}{}_\circ^{\,\,[I]}  \right\} \,,\label{J_adm_KS} 
\end{align}
where $\vec{v}{}^{\,\,[I]}$ and $\vec{d}{}^{\,\,[I]}$ are the velocity and position vectors of the individual black holes with respect to the background inertial frame, and $\vec{s}{}_\circ^{\,\,[I]}$ denotes the unit vectors pointing to directions of the individual spin vectors.

\section{The initial-boundary value problem}\label{initial-boundary}
\setcounter{equation}{0}

In advance of determining the ADM quantities relevant for asymptotically flat multiple black hole initial data configurations, one has to choose free data for the underlying initial-boundary problem. As a preparation for the asymptotic case first, by a straightforward adaptation of the method applied in \cite{racz_bh_ID}, considerations will be restricted to finite cubical domains.  

\medskip

\begin{figure}[htb]
	\begin{center}
	\psfragfig[width=9.3cm]{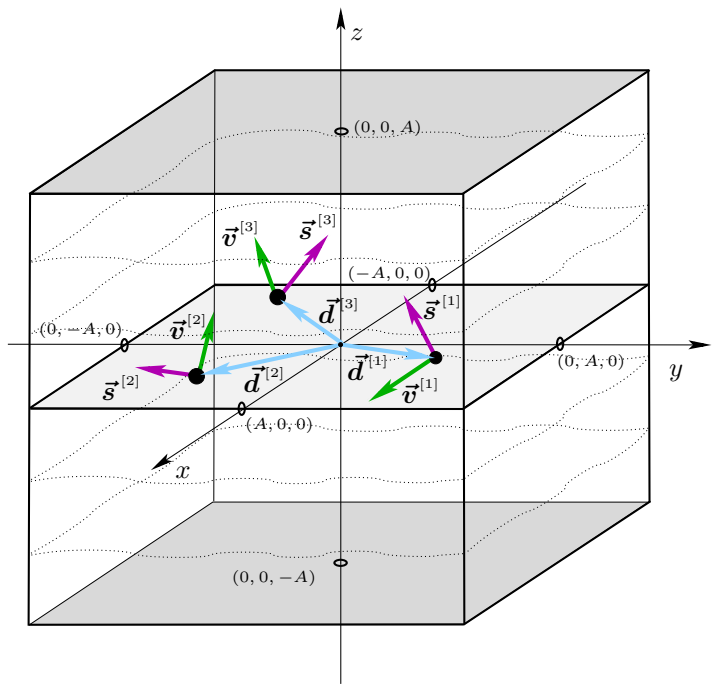}
	\end{center}
	\vskip-.5cm\caption{\footnotesize{ The initial data surface $\Sigma$, with a triple black hole system, is chosen to be the cube centered at the origin in $\mathbb{R}^3$ with edges $2A$. The initial data, to the system (\ref{bern_pde})--(\ref{ort_const_n}), is supposed to be specified on the horizontal squares, at $z=\pm A$, bounding the cube from above and below, whereas boundary values have to be given on the complementary part of the boundary comprised by four vertical squares.}}
	\label{kocka}
\end{figure}

Accordingly, the initial data surface $\Sigma$ is chosen to be a cube, centered at the origin in $\mathbb{R}^3$ (see Fig.\,\ref{kocka}) with boundary comprised by six squares each with edges $2 A$. 
By choosing the value of $A$ sufficiently large all the individual black holes will be contained in this cubical domain with suitable margin. The parabolic-hyperbolic system (\ref{bern_pde})--(\ref{ort_const_n}) has to be solved then as an initial-boundary value problem to which (local) well-posedness is guaranteed (see, e.g.~\cite{kreissl}) in those subregions of $\Sigma$ where (\ref{bern_pde}) is uniformly parabolic.

\medskip

As in \cite{racz_bh_ID}, the ring singularities of the individual black holes are assumed to be located momentarily on the $z=0$ plane in $\mathbb{R}^3$, and a foliation of $\Sigma$ by $z=const$ level surfaces will be applied. By an argument, analogous to the one applied in  \cite{racz_bh_ID}, the principal coefficient $\instar{K}$ of the parabolic equation (\ref{bern_pde}) can be shown to vanish on the  $z=0$ plane  dividing $\Sigma$ into two disjoint subsets. Accordingly, (\ref{bern_pde}) is uniformly parabolic on the disjunct subregions of $\Sigma$ located above and below of the $z=0$ surface, and one could also investigate the well-posedness of the parabolic-hyperbolic system (\ref{bern_pde})--(\ref{ort_const_n}) in these subregions. Nevertheless, as indicated in the introduction, instead of attempting to do so we shall simply assume that the global existence and uniqueness of solutions, along with their proper matching, is guaranteed. Note that the pertinent initial values are supposed to be specified on the horizontal $z=\pm A$ squares, whereas the boundary values have to be given on the four vertical sides of the cube (see Fig.\,\ref{kocka}).

\subsection{The asymptotic properties of the initial-boundary data}

The model of multiple black hole configurations, as introduced in the previous section, makes use of finite cubical domains. Therefore, to investigate the asymptotic properties of the corresponding initial-boundary data configurations, one has to consider sequences of solutions to the initial-boundary value problem such that the edges tend to infinity.
The individual members of such a sequence will differ slightly from global (up to spacelike infinity) solutions to the constraints.
Nevertheless, due to the asymptotic flatness of the auxiliary metric \eqref{eq:bks4}, the deviations are expected to be smaller and smaller as the boundary is pushed further and further towards spacelike infinity.

\medskip

Next we fix the freely specifiable fields $\widehat N^i,\widehat \gamma_{ij}, \boldsymbol\kappa$ and $\interior{\rm\bf K}_{ij}$ to coincide with the auxiliary fields $ \preA{\widehat N}^i , \preA{\widehat \gamma}_{ij}, \preA{\boldsymbol\kappa}$ and $\preA{\interior{\rm{\bf K}}}_{ij}$, respectively. The proper fall off property of this part of the data is guaranteed by the fact that (\ref{eq:bks4}) is asymptotically flat. As indicated in the previous sections, the initialization of the constrained variables $\widehat N, {\rm\bf K}^l{}_{l}$ and ${\rm\bf k}{}_{i}$ also happens by utilizing the auxiliary metric (\ref{eq:bks4}).

\medskip

In proceeding let us sum up what we already have by hand. Taking into account that the solution is assumed to be asymptotically flat and the Regge-Teitelboim parity conditions also hold, the fields $\widehat N, {\rm\bf K}^l{}_{l}$ and ${\rm\bf k}{}_{i}$ are expected to satisfy
\begin{equation} \label{constr_af}
	\big|\partial^{\alpha}\big(\widehat{N}-1\big)(\vec{x})\big|=\mathcal{O}(|\vec{x}|^{-|\alpha|-1})\,, \ 
	\big|\partial^{\beta}{\rm\bf K}^l{}_{l}(\vec{x})\big|=\mathcal{O}(|\vec{x}|^{-|\beta|-2})\,, \ 
	\big|\partial^{\beta}{\rm\bf k}{}_{i}(\vec{x})\big|=\mathcal{O}(|\vec{x}|^{-|\beta|-2})\,,
\end{equation}
and the parity conditions
\begin{align} 
 \big|\,\partial^{\alpha}\big(\,\widehat{N}(\vec{x}) - \widehat{N}(-\vec{x}) \,\big)\,\big| = {} &\mathcal{O}(\,|\vec{x}|^{-|\alpha|-2}\,)
\label{fa_Nhat} \\
\big|\,\partial^{\beta}\big(\,{\rm\bf K}^l{}_{l}(\vec{x}) - {\rm\bf K}^l{}_{l}(-\vec{x})\,\big)\,\big| = {} & \mathcal{O}(\,|\vec{x}|^{-|\beta|-3}\,)
\label{fa_bfK}  \\ 
\big|\,\partial^{\beta} \big(\,{\rm\bf k}{}_{i}(\vec{x}) -{\rm\bf k}{}_{i}(-\vec{x})\,\big)\,\big|= {} & \mathcal{O}(\,|\vec{x}|^{-|\beta|-3}\,) \,. 
\label{fa_bfk}
\end{align}

\medskip

Since all the ADM quantities are well-defined for the superposed Kerr-Schild configurations, the auxiliary fields $\preA\widehat N, \preA{\rm\bf K}^l{}_{l}$ and $\preA{\rm\bf k}{}_{i}$ do also satisfy conditions analogous to (\ref{constr_af})-(\ref{fa_bfk}). 

\medskip

Consider now the asymptotic expansions of the constrained fields $\widehat N, {\rm\bf K}^l{}_{l}$ and ${\rm\bf k}{}_{i}$ and those of the corresponding auxiliary fields $\preA{\widehat N}, \preA{\rm\bf K}^l{}_{l}$ and $\preA{\rm\bf k}{}_{i}$ given in terms of various powers of $1/|\vec{x}|^{n}$, with integer $n>0$. Referring to what we have inferred concerning sequences of solutions defined on finite cubical domains, it is plausible to assume that as the boundaries of these domains are pushed further and further towards spacelike infinity higher than the leading order terms of the asymptotic expansion play less and less important role. 
Therefore, as the only sensible asymptotic behavior of the fields $\widehat N, {\rm\bf K}^l{}_{l}$ and ${\rm\bf k}{}_{i}$ associated with the pertinent asymptotically flat solution, the leading order terms in their asymptotic expansions are assumed to be equal to those of $\preA{\widehat N}, \preA{\rm\bf K}^l{}_{l}$ and $\preA{\rm\bf k}{}_{i}$, respectively. Obviously, all the higher order contributions may---and, in general, do indeed---differ from each other. Nevertheless, as the fields $\widehat N, {\rm\bf K}^l{}_{l}$, ${\rm\bf k}{}_{i}$ and $\preA{\widehat N}, \preA{\rm\bf K}^l{}_{l}$, $\preA{\rm\bf k}{}_{i}$ are assumed to agree at leading order, respectively, their deviations\,\footnote{These deviations will be denoted by $\Delta$ followed by the pertinent quantities in square brackets, as they are spelled out explicitly in the first two terms of (\ref{Delta_Nhat})--(\ref{Delta_bfk}).} are expected to satisfy the following relations
\begin{align} 
	\big|\,\partial^{\alpha}\Delta[\widehat{N}](\vec{x})\,\big|= {} & \big|\,\partial^{\alpha}\big(\,\widehat{N}- \preA{\widehat{N}}\,\big)(\vec{x})\,\big|=\mathcal{O}(\,|\vec{x}|^{-|\alpha|-2}\,)
	\label{Delta_Nhat} \\
	\big|\,\partial^{\beta}\Delta[{\rm\bf K}^l{}_{l}](\vec{x})\,\big|= {} & \big|\,\partial^{\beta}\big(\,{\rm\bf K}^l{}_{l}-\preA{\rm\bf K}^l{}_{l}\,\big)(\vec{x})\,\big|= \mathcal{O}(\,|\vec{x}|^{-|\beta|-3}\,)
	\label{Delta_bfK}  \\ 
	\big|\,\partial^{\beta}\Delta[{\rm\bf k}{}_{i}](\vec{x})\,\big|= {} & \big|\,\partial^{\beta} \big(\,{\rm\bf k}{}_{i}-\preA{\rm\bf k}{}_{i}\,\big)(\vec{x})\,\big|= \mathcal{O}(\,|\vec{x}|^{-|\beta|-3}\,)
	\label{Delta_bfk}
\end{align}
and
\begin{align} 
\big|\,\partial^{\alpha}\big\{\Delta[\widehat{N}](\vec{x})-\Delta[\widehat{N}](-\vec{x})\big\}\,\big|= {} & \mathcal{O}(\,|\vec{x}|^{-|\alpha|-3}\,)
\label{ref_Delta_Nhat} \\
\big|\,\partial^{\beta}\big\{\Delta[{\rm\bf K}^l{}_{l}](\vec{x})-\Delta[{\rm\bf K}^l{}_{l}](-\vec{x})\big\}\,\big|= {} &  \mathcal{O}(\,|\vec{x}|^{-|\beta|-4}\,)
\label{ref_Delta_bfK}  \\ 
\big|\,\partial^{\beta}\big\{\Delta[{\rm\bf k}{}_{i}](\vec{x})-\Delta[{\rm\bf k}{}_{i}](-\vec{x})\big\}\,\big|= {} &  \mathcal{O}(\,|\vec{x}|^{-|\beta|-4}\,)\,.
\label{ref_Delta_bfk}
\end{align}

\section{The determination of the ADM quantities}
\label{adm_quantities}
\setcounter{equation}{0}

Now we are in the position to compare the ADM quantities of the superposed Kerr-Schild metric with those of the corresponding multiple black hole initial data. In particular, we shall show that  conditions (\ref{Delta_Nhat})-(\ref{ref_Delta_bfk}), along with the choices we made for the other auxiliary variables, guarantee that the two sets of ADM quantities are pairwise equal to each other. 
In the following subsections a case by case verification of this claim will be provided.

\subsection{The ADM mass}
\label{adm_mass}

Consider first the ADM mass. Start by replacing the flux integral applied in (\ref{int:Madm}) by a slightly different flux integral
\begin{equation}\label{sqMadm}
	M^{\,ADM}= \frac{1}{16\pi}\,\gsqint_\infty 
	\left[\, \partial_i h_{ij} - \partial_j h_{ii} \,\right] n^j  {\rm d}C  \,,
\end{equation} 
where the symbol $\sqint_\infty$ denotes the limit of integrals over the boundary of a sequence of co-centered cubes while the length of their edges tend to infinity, whereas $n^i$ and $ {\rm d}C$ denote the outward pointing unit normal vector and the volume element on the squares bounding individual cubes in this sequence. 

\bigskip

At the first glance it may not be obvious that the flux integrals over concentric spheres can be replaced by flux integrals over boundaries of co-centered cubical regions. Note, however, that to any individual member of these cubes there always exist a minimal radius sphere that contains the cube, and a maximal radius sphere that is contained by the cube. Clearly, either the minimal or maximal radius of spheres are applied to construct a sequence, the flux integrals defined with respect to them tend to the ADM mass. Thereby, the flux integrals evaluated on the boundaries of the cubes have to tend to the ADM mass as well. Accordingly, the limits of the integrals in (\ref{int:Madm}) and (\ref{int:sqMadm}) have to be equal to each other.

\bigskip

Returning to the main line of the argument note that our aim here is to show that the difference $\Delta[M^{\,ADM}]=M^{\,ADM}-\preA M^{\,ADM}$ between the ADM mass of the physical solution and that of the superposed Kerr-Schild black holes is zero. To see that this is indeed the case, note first that by virtue of  (\ref{sqMadm}) 
\begin{equation}\label{int:sqMadm}
\Delta[M^{\,ADM}]= \frac{1}{16\pi}\,\gsqint_\infty 
\left[\, \partial_i\,( h_{ij}-\preA h_{ij})  - \partial_j\,( h_{ii}-\preA h_{ii}) \,\right] n^j\,  {\rm d}C \,.
\end{equation}
Evaluating the integrands we need to determine first the involved derivatives. In doing so note that the difference $h_{ij}-\preA\hskip-0.4mm h_{ij}$, in virtue of the relations 
\begin{equation}\label{spec_rel}
	\widehat \gamma_{ij}= \preA{\widehat \gamma_{ij}}, \ \ \widehat  n_i={\widehat N} \,\delta_{iz}\ \ {\rm and} \ \ \preA{\widehat  n}_i=\preA{\widehat N}\,\delta_{iz}\,,
\end{equation}
reads as
\begin{equation}\label{delta:eq:bks3}
h_{ij}-\preA\hskip-0.4mm h_{ij}= \widehat  n_i\, \widehat n_j - \preA{\widehat  n}_i \preA{\widehat n}_j = \delta_{iz}\delta_{jz}\,({\widehat N}^2-\preA\hskip-0.4mm{\widehat N}^2 ) = \delta_{iz}\delta_{jz}\,\Delta[{\widehat N}^2] \,.
\end{equation} 
(\ref{delta:eq:bks3}) implies then  that 
\begin{equation}\label{partial1:eq:bks3}
\partial_k\,( h_{ij}-\preA h_{ij}) = \delta_{iz}\delta_{jz}\partial_k\,(\Delta[{\widehat N}^2]) \,,
\end{equation}
and also that 
\begin{equation}\label{partial2:eq:bks3}
\partial_i\,( h_{ij}-\preA\hskip-0.4mm h_{ij}) = \delta_{jz}\partial_z\,(\Delta[{\widehat N}^2]) \quad{\rm and}\quad \partial_j\,( h_{ii}-\preA\hskip-0.4mm h_{ii}) = \partial_j\,(\Delta[{\widehat N}^2])\,.
\end{equation}

\medskip

By combining (\ref{sqMadm}), (\ref{int:sqMadm}) and (\ref{partial2:eq:bks3}) we get then that 
\begin{align}\label{int:sqMadm2}
\Delta[M^{\,ADM}]= {} &  \frac{1}{16\pi}\,\gsqint_\infty 
\left[\, \delta_{jz}\partial_z\,(\Delta[{\widehat N}^2])  - \partial_j\,(\Delta[{\widehat N}^2]) \,\right] n^j \, {\rm d}C \nonumber \\
= {} & -\frac{1}{16\pi}\,\gsqint_\infty 
\left[\,  \partial_x\,(\Delta[{\widehat N}^2])\,  \vec{n}{}^{\,x}_{\pm} + \partial_y\,(\Delta[{\widehat N}^2]) \, \vec{n}{}^{\,y}_{\pm}\,\right]  {\rm d}C \,,
\end{align}
where $\vec{n}{}^{\,x}_{\pm}$ and $\vec{n}{}^{\,y}_{\pm}$ denote the outward pointing unit normal vectors to the  squares in the $x=\pm A$ and $y=\pm A$ plains, bounding the cubical region on Fig.\,\ref{kocka}. Since $\vec{n}{}^{\,x}_{-} =-\vec{n}{}^{\,x}_{+}$ and $\vec{n}{}^{\,y}_{-} =-\vec{n}{}^{\,y}_{+}$, the two terms given explicitly in the integrands stand indeed for four terms. 

\medskip

By applying then the replacements $y\rightarrow x\,\upsilon$ and $z\rightarrow x\,\zeta$ in the first term of (\ref{int:sqMadm2})  evaluated on the squares in the $x=\pm A$ plains, and also the replacements  $x\rightarrow y\,\xi$ and $z\rightarrow y\,\zeta$ in the second term of (\ref{int:sqMadm2}) evaluated on the squares in the $y=\pm A$ plains we get 
\begin{align}\label{int:sqMadm3}
\Delta[M^{\,ADM}]= {} & -\frac{1}{16\pi}\left[\,\gsqint_{-1,-1}^{1,1} \lim\limits_{{x}\,\rightarrow \infty}  \left(\, x^2\,\partial_x\,(\Delta[{\widehat N}^2])\,  n^x_{\pm}\,\right)\, {\rm d}\upsilon\, {\rm d}\zeta \right. \nonumber \\   {} &  \hskip3cm + \left. 
\gsqint_{-1,-1}^{1,1} \lim\limits_{{y}\,\rightarrow \infty}  \left(\, y^2\,\partial_y\,(\Delta[{\widehat N}^2])\,  n^y_{\pm}\,\right)\,\, {\rm d}\xi \,{\rm d}\zeta \,\right] 
\,,
\end{align}
where  ${n}{}^{\,x}_{\pm}$ and ${n}{}^{\,y}_{\pm}$ are scalars taking the values ${n}{}^{\,x}_{\pm}=\pm 1$ and ${n}{}^{\,y}_{\pm}=\pm 1$ on the squares in the $x=\pm A$ and $y=\pm A$ plains, respectively.  
%
%
Note that the replacements applied in the above integral transformations are analogous to the ones used in case of sequences of spheres, where the integrals formally are given over a unit sphere with angular coordinates $\theta$ and $\phi$ ranging through their usual intervals. In both cases with the help of these integral transformations the limits of integrals  can be evaluated by inspecting the limits of the yielded integrands. 

\medskip
  
Using then
\begin{equation}\label{hatNsq}
\Delta[{\widehat N}^2]=(\Delta[{\widehat N}])^2+2\,\preA\hskip-0.5mm{\widehat N}\, \Delta[{\widehat N}]\,, 
\end{equation}
along with (\ref{Delta_Nhat}), applied for $\Delta[{\widehat N}]$, and (\ref{fa_Nhat}), applied for $\preA{\widehat N}$, respectively, we get that each of the terms $x_i\,\Delta[{\widehat N}]$,  $x_i^2\,(\partial_{x_i}\Delta[{\widehat N}])$ and $x_i\,(\partial_{x_i}\preA{\widehat N})$ appearing in the integrand of (\ref{int:sqMadm3}) is at most of order $\mathcal{O}(|\vec{x}|^{-1})$. Here the relations  $x_i=({x_i}/{|\vec{x}|})\,{|\vec{x}|}=\interior{x}_i\,{|\vec{x}|}$ and $|\interior{x}_i|\leq 1$ were also used. This, in turn, implies that the limits exist and they all vanish which verifies that $\Delta[M^{\,ADM}]=0$. 

\subsection{The center of mass}
\label{center_mass}

Rephrasing (\ref{center_mass}), by using integrals over cubical domains, we get that the center of mass can be given by the flux integral
\begin{equation}\label{sqcenter}
M^{\,ADM}d_l =  \frac{1}{16\pi}\,\gsqint_\infty 
\Big\{ x_l \left[\, \partial_k h_{kj} - \partial_j h_{kk} \,\right]   -  \left[\, h_{kj}\,\delta^k{}_l -  h_{kk}\,\delta_{lj} \,\right] \Big\}\,n^j  
{\rm d}C\,. 
\end{equation}

\medskip

Taking now into account the consequences of (\ref{delta:eq:bks3}) we get
\begin{equation}\label{Delta:eq:bks3}
\Delta[\,h_{kj}\,] = \delta_{kz}\delta_{jz}\,\Delta[{\widehat N}^2] \quad{\rm and}\quad \Delta[\,h_{kk}\,] = \Delta[{\widehat N}^2]\,.
\end{equation}
A straightforward calculation, consisting of steps analogous to the ones applied in the previous subsection, yields then
\begin{align}\label{int:center}
	\Delta[M^{\,ADM}d_l]= {} &  \frac{1}{16\pi}\,\gsqint_\infty 
	\Big\{ x_l \left[\, \partial_k ( h_{kj}-\preA h_{kj}) - \partial_j ( h_{kk}-\preA h_{kk}) \,\right] \Big. \nonumber \\ {} & \hskip 2cm - \Big. \left[\, ( h_{kj}-\preA h_{kj})\,\delta^k{}_l -  ( h_{kk}-\preA h_{kk}) \,\delta_{lj} \,\right] \Big\}\,n^j  {\rm d}C 
	\nonumber \\ {} & \hskip-1.6cm = \frac{1}{16\pi}\,\gsqint_\infty 
	\Big\{ x_l \left[\, \partial_k (\Delta[h_{kj}]) - \partial_j ( \Delta[h_{kk}]) \,\right] 
	- \left[\,\Delta[h_{kj}] \,\delta^k{}_l - \Delta[h_{kk}] \,\delta_{lj} \,\right] \Big\}\,n^j  {\rm d}C \nonumber \\ 
	{} & \hskip-1.6cm 	=  \frac{1}{16\pi}\,\gsqint_\infty 
\left\{\, -x_l\left[\,\partial_x\,(\Delta[{\widehat N}^2])\,\vec{n}^{\,x}_{\pm} + \partial_y\,(\Delta[{\widehat N}^2])\,\vec{n}^{\,y}_{\pm} \,\right] \right. \nonumber\\ {} & \hskip 4.8cm + \left.
\left[\,\delta_{lx}\,  \vec{n}^{\,x}_{\pm}+\delta_{ly}\,  \vec{n}^{\,y}_{\pm} \right]  \,\Delta[{\widehat N}^2] \,\right\} {\rm d}C \nonumber \\ 
{} & \hskip-1.6cm =  -\frac{1}{16\pi}\,\gsqint_\infty 
\left\{\left[\, x_l\,\partial_x(\Delta[{\widehat N}^2]) - \delta_{lx}\,\Delta[{\widehat N}^2]\,\right] \vec{n}^{\,x}_{\pm}  \right. \nonumber\\ {} & \hskip 4.5cm + \left.\left[\, x_l\,\partial_y(\Delta[{\widehat N}^2]) - \delta_{ly}\,\Delta[{\widehat N}^2]\,\right] \vec{n}^{\,y}_{\pm}   \,\right\} {\rm d}C \,.
\end{align}

Accordingly, for the $x$-component of the deviation $\Delta[M^{\,ADM}d_l]$ 
\begin{align}\label{int:center_x}
\hskip-0.3cm\Delta[M^{\,ADM}d_x]= {} &  -\frac{1}{16\pi}\,\gsqint_\infty 
\left\{\left[\, x\,\partial_x(\Delta[{\widehat N}^2]) - \Delta[{\widehat N}^2]\,\right] \vec{n}^{\,x}_{\pm}  + \left[\, x\,\partial_y(\Delta[{\widehat N}^2])]\,\right] \vec{n}^{\,y}_{\pm}   \,\right\} {\rm d}C \nonumber\\
= {} &  -\frac{1}{16\pi}\,\left[\,\,\gsqint_{-1,-1}^{1,1} \lim\limits_{{x}\,\rightarrow \infty}  \left\{\,\left[ x^3\,\partial_x(\Delta[{\widehat N}^2])-x^2\,\Delta[{\widehat N}^2]\,\right]  n^x_{\pm}\,\right\}\, {\rm d}\upsilon\, {\rm d}\zeta \right. \nonumber
\\{} & \left. \hskip3.cm + \,\gsqint_{-1,-1}^{1,1} \lim\limits_{{y}\,\rightarrow \infty}  \left\{\,\left[ y^3\,\partial_y(\Delta[{\widehat N}^2])\,\right]  n^y_{\pm}\,\right\} \,\xi\, {\rm d}\xi \,{\rm d}\zeta \,\right]
\end{align}
hold.

\medskip

Taking into account (\ref{hatNsq}), along with the relations $n^x_{+}=-n^x_{-}=1$ and $n^y_{+}=-n^y_{-}=1$, in virtue of (\ref{fa_Nhat}), (\ref{Delta_Nhat}) and the Regge-Teitelboim condition (\ref{ref_Delta_Nhat}), it follows that either of the terms $x^2\Delta[{\widehat N}^2]$,  $x^3(\partial_{x}\Delta[{\widehat N}^2])$,  $y^3(\partial_{y}\Delta[{\widehat N}^2])$ and $x_i^2\,(\partial_{x_i}\preA{\widehat N})$ in the integrands, is at most of order $\mathcal{O}(|\vec{x}|^{-1})$. This, in turn, implies, as above, that the pertinent limits exist and they vanish as we intended to show. 

\medskip

In virtue of the relations,
\begin{align}\label{int:center_y}
\hskip-0.5cm\Delta[M^{\,ADM}d_y] 
= {} &  -\frac{1}{16\pi}\,\left[\,\,\gsqint_{-1,-1}^{1,1} \lim\limits_{{x}\,\rightarrow \infty} \left\{\,\left[ x^3\,\partial_x(\Delta[{\widehat N}^2])\,\right]  n^x_{\pm}\,\right\} \, \upsilon \, {\rm d}\upsilon\, {\rm d}\zeta \right. \nonumber
\\{} & \left. \hskip1.5cm + \,\gsqint_{-1,-1}^{1,1} \lim\limits_{{y}\,\rightarrow \infty}  \left\{\,\left[ y^3\,\partial_y(\Delta[{\widehat N}^2])-y^2\,\Delta[{\widehat N}^2]\,\right]  n^y_{\pm}\,\right\} \, {\rm d}\xi \,{\rm d}\zeta \,\right]
\end{align}
and
\begin{align}\label{int:center_z}
\hskip-0.5cm\Delta[M^{\,ADM}d_z] 
= {} &  -\frac{1}{16\pi}\,\left[\,\,\gsqint_{-1,-1}^{1,1} \lim\limits_{{x}\,\rightarrow \infty} \left\{\,x^3\,\partial_x(\Delta[{\widehat N}^2])\,  n^x_{\pm}\,\right\} \, \zeta \, {\rm d}\upsilon\, {\rm d}\zeta \right. \nonumber
\\{} & \left. \hskip2.5cm + \,\gsqint_{-1,-1}^{1,1} \lim\limits_{{y}\,\rightarrow \infty}  \left\{\,y^3\,\partial_y(\Delta[{\widehat N}^2])\, n^y_{\pm}\,\right\}\, \zeta \, {\rm d}\xi \,{\rm d}\zeta \,\right]\,,	
\end{align}
arguments, analogous to the one applied above, can be used to show the vanishing of the $y$- and  $z$-components of the deviation $\Delta[M^{\,ADM}d_l]$.

\subsection{The linear momentum}
\label{linear_momentu}

Consider now the linear momentum determined by the flux integral 
\begin{equation}\label{P_ADM_sq}
{P}_i^{\,ADM} = \frac{1}{8\pi}\,\gsqint_\infty  
\left[\, K_{ij} - h_{kj}\,K^{l}{}_{l} \,\right] n^j  {\rm d}C \,.
\end{equation}

In verifying that the individual components of $\Delta[{P}_i^{\,ADM}]$ vanish, respectively, it is rewarding to rephrase first the term $K_{ij}-h_{ij}\,{K}^l{}_{l}$ in (\ref{P_ADM_sq}) in terms of the new variables we introduced. In doing so we get, in virtue of (\ref{hij})--(\ref{trace_free}), that 
\begin{equation}\label{decopKij_new}
K_{ij}= \boldsymbol\kappa \,\widehat n_i \widehat n_j  + \left[\widehat n_i \,{\rm\bf k}{}_j  
+ \widehat n_j\,{\rm\bf k}{}_i\right]  + \left(\interior{\rm\bf K}_{ij}+\tfrac12\,\widehat \gamma_{ij}\,{\rm\bf K}^l{}_{l}\right)\,,
\end{equation}
and
\begin{equation}\label{decoptrKij}
{K}^l{}_{l}=\boldsymbol\kappa + {\rm\bf K}^l{}_{l}\,.
\end{equation}

\medskip

It is straightforward to see then that 
\begin{equation}\label{integrand_P}
K_{ij}-h_{ij}\,{K}^l{}_{l}= \left[\widehat n_i \,{\rm\bf k}{}_j  
+ \widehat n_j\,{\rm\bf k}{}_i\right]  + \left(\interior{\rm\bf K}_{ij}+\tfrac12\,\widehat \gamma_{ij}\,{\rm\bf K}^l{}_{l}\right) -\left\{
\widehat \gamma_{ij}\,\boldsymbol\kappa + (\widehat \gamma_{ij} + \widehat n_i \widehat n_j )\,{\rm\bf K}^l{}_{l}
\right\}\,,
\end{equation}
and, by applying (\ref{spec_rel}), that
\begin{equation}\label{ref_integrand_P}
\Delta[K_{ij}-h_{ij}\,{K}^l{}_{l}]= \delta_{iz}\, \Delta[\widehat{N}\,{\rm\bf k}{}_j] + \delta_{jz}\, \Delta[\widehat{N}\,{\rm\bf k}{}_i]  
 - \tfrac12\,\widehat \gamma_{ij}\,\Delta[{\rm\bf K}^l{}_{l}] - \delta_{iz}\delta_{jz}\,\Delta[\widehat{N}^2\,{\rm\bf K}^l{}_{l}]\,.
\end{equation}

It follows then that for the individual components of $\Delta[{P}_i^{\,ADM}]$ the relations 
\begin{align}
\Delta[{P}_x^{\,ADM}]={} & \frac{1}{8\pi}\,\gsqint_\infty 
\left\{-\tfrac12\,\widehat \gamma_{xx}\,\Delta[{\rm\bf K}^l{}_{l}]\,  \vec{n}^x_{\pm} -\tfrac12\,\widehat \gamma_{xy}\,\Delta[{\rm\bf K}^l{}_{l}]\,  \vec{n}^y_{\pm}  + \Delta[\widehat{N}\,{\rm\bf k}{}_x]\,  \vec{n}^z_{\pm} \right\} {\rm d}C \label{rel_P_x} \\ 
\Delta[{P}_y^{\,ADM}]={} & \frac{1}{8\pi}\,\gsqint_\infty 
\left\{-\tfrac12\,\widehat \gamma_{yx}\,\Delta[{\rm\bf K}^l{}_{l}]\,  \vec{n}^x_{\pm} -\tfrac12\,\widehat \gamma_{yy}\,\Delta[{\rm\bf K}^l{}_{l}]\,  \vec{n}^y_{\pm}  + \Delta[\widehat{N}\,{\rm\bf k}{}_y]\,  \vec{n}^z_{\pm} \right\} {\rm d}C \label{rel_P_y} \\ 
\Delta[{P}_z^{\,ADM}]={} & \frac{1}{8\pi}\,\gsqint_\infty 
\left\{\,\Delta[\widehat{N}\,{\rm\bf k}{}_x]\,\vec{n}^x_{\pm} + \Delta[\widehat{N}\,{\rm\bf k}{}_y]\,  \vec{n}^y_{\pm}  - \Delta[\widehat{N}^2\,{\rm\bf K}^l{}_{l}] \,  \vec{n}^z_{\pm} \right\} {\rm d}C \label{rel_P_z} 
\end{align}
hold, where ${\rm\bf k}{}_i\,\widehat{n}{}^i=0$ and $\widehat \gamma_{ij}\,\widehat{n}{}^i=0$ had also been used.

\medskip

In verifying that each of the components of $\Delta[\vec{P}^{\,ADM}]$ vanish---besides the replacements $y\rightarrow x\,\upsilon$ $z\rightarrow x\,\zeta$, and $x\rightarrow y\,\xi$ $z\rightarrow y\,\zeta$ on the $x=\pm A$ and $y=\pm A$ surfaces, respectively---the transformations $x\rightarrow z\,\xi$ and $y\rightarrow z\,\upsilon$ have also to be performed on the $z=\pm A$ surfaces since the integrands there, as opposed to the previous two cases, are not identically zero. By applying the corresponding integral transformations we get from (\ref{rel_P_x}) that
\begin{align}
\Delta[{P}_x^{\,ADM}]= -\frac{1}{16\pi}{} & \left[\,\gsqint_{-1,-1}^{1,1} 
\lim\limits_{{x}\,\rightarrow \infty}\left(\,\widehat \gamma_{xx}\,n^x_{\pm} \left\{x^2\Delta[{\rm\bf K}^l{}_{l}] \right\}\, \right)\,{\rm d}\upsilon\,{\rm d}\zeta \right.\\
\,{} &  \, +\left.\,\gsqint_{-1,-1}^{1,1} \lim\limits_{{y}\,\rightarrow \infty}\left(\,
\widehat \gamma_{xy}\,n^y_{\pm} \left\{y^2\Delta[{\rm\bf K}^l{}_{l}] \right\}\, \right)\,{\rm d}\xi\,{\rm d}\zeta \nonumber \right. \\
{} & \hskip-.3cm -2\left. \,\gsqint_{-1,-1}^{1,1} \lim\limits_{{z}\,\rightarrow \infty}\left(\,
n^z_{\pm}\left[\left\{ z^2 \,\Delta[\widehat{N}]\,\preA{\rm\bf k}{}_x\right\} + \preA\widehat{N}\left\{z^2 \,\Delta[{\rm\bf k}{}_x]  \right\}\,\right]\, \right)\,{\rm d}\xi\,{\rm d}\upsilon\,\right]\,. \nonumber
\end{align}
Taking then into account the fall off conditions in (\ref{Delta_Nhat}) and (\ref{Delta_bfk}), along with the one satisfied, in virtue of (\ref{constr_af}), by $\preA{\rm\bf k}{}_x$, we get that all the terms in braces are at most of order $\mathcal{O}(|\vec{x}|^{-1})$ which verifies that $\Delta[{P}_x^{\,ADM}]$ vanishes.

\medskip

As the terms in (\ref{rel_P_y}) are similar to those in   (\ref{rel_P_x}), a completely analogous argument applies to the $y$-component of $\Delta[{P}_i^{\,ADM}]$. 

\medskip

To show that $z$-component, $\Delta[{P}_z^{\,ADM}]$, does also vanish note first that (\ref{rel_P_z}) can be given as 
\begin{align}
\Delta[{P}_z^{\,ADM}]= \frac{1}{8\pi} {} & {} \left[\,\gsqint_{-1,-1}^{1,1} \lim\limits_{{x}\,\rightarrow \infty}\left(\,
n^x_{\pm}\left[\left\{ x^2 \,\Delta[\widehat{N}]\,\preA{\rm\bf k}{}_x\right\} + \preA\widehat{N}\left\{x^2 \,\Delta[{\rm\bf k}{}_x]  \right\}\,\right]\, \right)
\,{\rm d}\upsilon\,{\rm d}\zeta \right.
\nonumber \\
{} & +  \left.\,\gsqint_{-1,-1}^{1,1} \lim\limits_{{y}\,\rightarrow \infty}\left(\,
n^y_{\pm}\left[\left\{ y^2 \,\Delta[\widehat{N}]\,\preA{\rm\bf k}{}_y\right\} + \preA\widehat{N}\left\{y^2 \,\Delta[{\rm\bf k}{}_y]  \right\}\,\right]\, \right)
\,{\rm d}\xi\,{\rm d}\zeta\right.
\nonumber \\
- {} & \left. \,\gsqint_{-1,-1}^{1,1} \lim\limits_{{z}\,\rightarrow \infty}\left(\,
n^z_{\pm}\left[\,\left\{ z^2 \,\Delta[\widehat{N}^2]\,\preA{\rm\bf K}^l{}_{l}\right\} + \preA\widehat{N}^2\left\{z^2 \,\Delta[{\rm\bf K}^l{}_{l}]  \right\}\,\right]\,\right)\,{\rm d}\xi\,{\rm d}\upsilon\,\right]\,.
\nonumber\\ 
\end{align}
Now, by taking into account the fall off conditions listed in relations (\ref{Delta_Nhat})--(\ref{Delta_bfk}), along with those satisfied, in virtue of (\ref{constr_af}), by $\preA{\rm\bf k}{}_x$, $\preA{\rm\bf k}{}_y$ and $\preA{\rm\bf K}^l{}_{l}$, we get again that all the terms in braces are at most of order $\mathcal{O}(|\vec{x}|^{-1})$ which verifies that $\Delta[{P}_z^{\,ADM}]$ vanishes as we desired to show.

\medskip

Note that, as expected, in verifying that the physical ADM mass and linear momentum are equal to the ADM mass and linear momentum of the superposed Kerr-Schild black holes, no use of the Regge-Teitelboim conditions had to be made.  

\subsection{The angular momentum}
\label{angular_momentu}

Consider, finally, the angular momentum determined by the flux integral
\begin{equation}\label{J_adm}
{J}_i^{\,ADM} = \frac{1}{8\pi} \,\,\gsqint_\infty 
\left[\, K_{kl} - h_{kl}\,K^{m}{}_{m} \,\right] \,\epsilon_{i}{}^{jk}x_j\, n^l  {\rm d}C \,.
\end{equation}
Thus we have
\begin{equation}\label{Delta_J_adm}
\Delta[{J}_i^{\,ADM}] = {J}_i^{\,ADM} - \preA{J}_i^{\,ADM} = \frac{1}{8\pi} \,\,\gsqint_\infty \Delta[\,{\widetilde J}_{il}\,]
\, n^l \, {\rm d}C \,,
\end{equation}
where $\Delta[{\widetilde J}_{il}]=\Delta\left[\, K_{kl} - h_{kl}\,K^{m}{}_{m} \,\right] \,\epsilon_{i}{}^{jk}x_j$.

By a direct calculation, consisting of steps analogous to the ones applied in the previous subsection in evaluating $\Delta[K_{ij}-h_{ij}\,{K}^l{}_{l}]$ in (\ref{ref_integrand_P}), the relations
\begin{align}
 \Delta[\,{\widetilde J}_{xl}\,] ={} & y \left[\,\Delta[\widehat{N}\,{\rm\bf k}{}_l] - \delta_{lz}\,\Delta[\widehat{N}^2\,{\rm\bf K}^m{}_{m}]\, \right]-
z \left[\,\delta_{lz}\,\Delta[\widehat{N}\,{\rm\bf k}{}_y] - \tfrac12\,\widehat\gamma_{ly}\,\Delta[{\rm\bf K}^m{}_{m}]\, \right]
\\ 
\Delta[\,{\widetilde J}_{yl}\,] ={} &   z \left[\,\delta_{lz}\,\Delta[\widehat{N}\,{\rm\bf k}{}_x] - \tfrac12\,\widehat\gamma_{lx}\,\Delta[{\rm\bf K}^m{}_{m}]\, \right]- x \left[\,\Delta[\widehat{N}\,{\rm\bf k}{}_l] - \delta_{lz}\,\Delta[\widehat{N}^2\,{\rm\bf K}^m{}_{m}] \,\right]
\\ 
\Delta[\,{\widetilde J}_{zl}\,] ={} &  x \left[\,\delta_{lz}\,\Delta[\widehat{N}\,{\rm\bf k}{}_y] - \tfrac12\,\widehat\gamma_{ly}\,\Delta[{\rm\bf K}^m{}_{m}]\, \right] - y \left[\,\delta_{lz}\,\Delta[\widehat{N}\,{\rm\bf k}{}_x] - \tfrac12\,\widehat\gamma_{lx}\,\Delta[{\rm\bf K}^m{}_{m}]\, \right]
\end{align}
can be seen to hold. 

\medskip

It is then straightforward to verify that 
\begin{align}
\Delta[{J}_x^{\,ADM}]= \frac{1}{8\pi}\,\,\gsqint_\infty {} & \left\{\,\left[\, y \,\Delta[\widehat{N}\,{\rm\bf k}{}_l] + \tfrac12\,z\,\widehat\gamma_{xy}\,\Delta[{\rm\bf K}^m{}_{m}]\,\right] \vec{n}^{\,x}_{\pm} \right.  \nonumber \\
{} & \left. \hskip1cm + \,\left[\, y \,\Delta[\widehat{N}\,{\rm\bf k}{}_y] + \tfrac12\,z\,\widehat\gamma_{yy}\,\Delta[{\rm\bf K}^m{}_{m}]\,\right] \vec{n}^{\,y}_{\pm} \right. \nonumber \\
{} & 
\left. \hskip2cm  - \left[\,y\,\Delta[\widehat{N}^2\,{\rm\bf K}^m{}_{m}]+z\,\Delta[\widehat{N}\,{\rm\bf k}{}_y] \,\right]  \vec{n}^{\,z}_{\pm} 
\,\right\}  {\rm d}C \label{J_x}
\end{align}
from which---by applying the integral transformations already used several times in the previous subsections, on the individual $x_i=\pm A$ plains---we get 
\begin{align}
\Delta[{J}_x^{\,ADM}]= \frac{1}{8\pi}{} &\left[\,\,\gsqint_{-1,-1}^{1,1} \lim\limits_{{x}\,\rightarrow \infty} \left\{ x^3 \left[\,\upsilon\, \,\Delta[\widehat{N}\,{\rm\bf k}{}_l] + \tfrac12\,\zeta\,\widehat\gamma_{xy}\,\Delta[{\rm\bf K}^m{}_{m}]\,\right] {n}^{\,x}_{\pm} \,\right\}\,{\rm d}\upsilon\,{\rm d}\zeta  \right.  \nonumber \\
{} & \left. \hskip0,3cm + \,\,\gsqint_{-1,-1}^{1,1} \lim\limits_{{y}\,\rightarrow \infty} \left\{ y^3 \left[\,\Delta[\widehat{N}\,{\rm\bf k}{}_y] + \tfrac12\,\zeta\,\widehat\gamma_{yy}\,\Delta[{\rm\bf K}^m{}_{m}]\,\right] {n}^{\,y}_{\pm}\,\right\}\,{\rm d}\xi\,{\rm d}\zeta  \right. \nonumber \\
{} & 
\left. \hskip.6cm  - \,\,\gsqint_{-1,-1}^{1,1} \lim\limits_{{z}\,\rightarrow \infty} \left\{ z^3 \left[\,\upsilon\,\Delta[\widehat{N}^2\,{\rm\bf K}^m{}_{m}]+ \Delta[\widehat{N}\,{\rm\bf k}{}_y] \,\right]  {n}^{\,z}_{\pm} \,\right\}\,{\rm d}\xi\,{\rm d}\upsilon 
\, \right] \,.
\end{align}

Taking now into account (\ref{hatNsq}), along with the relations $n^x_{+}=-n^x_{-}=1$ and $n^y_{+}=-n^y_{-}=1$, and also the boundedness of the components of $\widehat\gamma_{ij}$ and that of the coordinates $\xi,\upsilon, \zeta$, in virtue of (\ref{fa_Nhat})--(\ref{fa_bfk}), (\ref{Delta_Nhat})--(\ref{Delta_bfk}) and the Regge-Teitelboim condition  (\ref{ref_Delta_Nhat})--(\ref{ref_Delta_bfk}), it follows that either of the terms involved by the integrands, is at most of order $\mathcal{O}(|\vec{x}|^{-1})$ which implies the vanishing of $\Delta[{J}_x^{\,ADM}]$. 

\medskip

In virtue of the use of analogous terms in $\Delta[\,{\widetilde J}_{xl}\,]$ and $\Delta[\,{\widetilde J}_{yl}\,]$ a completely analogous argument can be seen to apply to the $y$-component of $\Delta[{J}_i^{\,ADM}]$. 

\medskip

Finally, $\Delta[{J}_z^{\,ADM}]$ can be evaluated as 
\begin{align}
\Delta[{J}_z^{\,ADM}]= -\frac{1}{16\pi}{} &\left[\,\,\gsqint_{-1,-1}^{1,1} \lim\limits_{{x}\,\rightarrow \infty} \Big\{ x^3 \big[\,\widehat\gamma_{xy} - \upsilon\,\widehat\gamma_{xx}\,\big]\,\Delta[{\rm\bf K}^m{}_{m}]\, {n}^{\,x}_{\pm} \,\Big\}\,{\rm d}\upsilon\,{\rm d}\zeta  \right.  \nonumber \\
{} & \left. \hskip0,3cm + \,\,\gsqint_{-1,-1}^{1,1} \lim\limits_{{y}\,\rightarrow \infty} \Big\{ y^3 \big[\,\xi\,\widehat\gamma_{yy} - \widehat\gamma_{yx}\,\big]\,\Delta[{\rm\bf K}^m{}_{m}]\, {n}^{\,y}_{\pm} \,\Big\}
\,{\rm d}\xi\,{\rm d}\zeta  \right. \nonumber \\
{} & 
\left. \hskip.6cm  - 2 \,\,\gsqint_{-1,-1}^{1,1} \lim\limits_{{z}\,\rightarrow \infty} \left\{ z^3 \left[\,\xi\,\Delta[\widehat{N}\,{\rm\bf k}{}_y]-\upsilon\,\Delta[\widehat{N}\,{\rm\bf k}{}_x] \,\right]  {n}^{\,z}_{\pm} \,\right\}\,{\rm d}\xi\,{\rm d}\upsilon 
\, \right] \,.
\end{align}
Then, by~making use of the fall off and Regge-Teitelboim conditions, (\ref{Delta_Nhat})--(\ref{ref_Delta_bfk}), and~by an argument that has already been applied several times above, the~vanishing of $\Delta[{J}_z^{\,ADM}]$ can be~inferred. 

\medskip

Putting all the results of the previous  subsections together we get the desired verification of Theorem\,\ref{theor}.

\section{Conclusions}\label{sec: conclusions}

Our primary aim was to answer the question raised in the title concerning the freedom we have in specifying the physical parameters of multiple black hole configurations. In doing so, a combination of the parabolic-hyperbolic formulation of constraints and superposed Kerr-Schild black holes were used. We treated only the case of multiple black hole systems where the ring singularities and the speeds of the individual Kerr-Schild black holes were confined to the $z=0$ plane. This, also meant that the spins were required to be aligned or anti-aligned to the $z$-axis. As there were no further restrictions on the input parameters, a significant number of multiple black hole configurations with immediate physical interest are covered by the investigated set.

\medskip

The main result of this paper can be formulated as follows:

\begin{theorem}\label{theor}\vskip-.2cm
	
	Suppose that an asymptotically flat solution to the initial-boundary value problem---deduced from the parabolic-hyperbolic form of the constrains, (\ref{bern_pde})--(\ref{ort_const_n})---exists such that the free data is chosen, as described in section\,\ref{initial-boundary}, by applying the superposed Kerr-Schild metric (\ref{eq:bks4}). Then, the ADM mass, center of mass, linear and angular momenta, relevant for the initialization of the corresponding multiple black hole system, can be given, as in  (\ref{adm_mass})--(\ref{angular_momentu}), in terms of the rest masses, positions, velocities, and spins of the involved individual Kerr-Schild black holes. In addition, all of these parameters can be prescribed in advance of solving the constraints.
	
\end{theorem}

Several remarkable features characterize the applied initial data construction. First, as we do not use conformal rescalings, our method retains the physically distinguished nature of $h_{ij}$ and $K_{ij}$. Second, the input parameters are the rest masses, the sizes, and orientations of the displacements, velocities, and spins of the individual black holes. Note that these are essentially the same as the input parameters of the post-Newtonian (PN) formalism. This provides significant interrelations between the PN and our fully relativistic setups. In particular, physically adequate choices of the orbital parameters could be made using the insights earned within the PN. Notably, as shown in this paper, each of the global ADM charges can also be given in terms of the input parameters, which is unprecedented in other methods to solve the constraints. More strikingly, the ADM mass, center of mass, linear and angular momenta of the binary system can be fixed in advance of solving the constraints.

\medskip

Despite the advantages discussed above, it is important to keep in mind that there is room for further investigations. There is an obvious interest to generalize the applied initial data construction to the case where the speed and spin vectors of involved individual Kerr-Schild black holes are arbitrarily pointing. Once this is done, there will also be a need to generalize the results covered by this paper.

\medskip

Another physically important generalization could be to replace asymptotic flatness with more realistic geometric assumptions. For instance, black holes could be placed in the environment of the expanding universe modeled by the standard Friedman-Lemaitre-Robertson-Walker solutions. Similarly, investigations of black hole systems in the McVittie background, studied, e.g., in \cite{Antoniou,nolan}, could also be of interest. Note that such a replacement will require a more substantial generalization of the construction as even the freely specifiable variables have to be altered significantly.

Both of the indicated problems---which certainly deserve further attention---are left open for future investigations.

\section*{Acknowledgments}

The author is grateful to Piotr Chru\'sciel,  Andor Frenkel, Jan Metzger, Christopher Nerz, Harald Pfeiffer, Bob Wald and Jeff Winicour for helpful comments. Thanks are due to the Albert Einstein Institute in Golm for its kind hospitality. This work was also supported in parts by the NKFIH grant K-115434. 


\end{document}